# Formation and propagation of matter wave soliton trains

Kevin E. Strecker, Guthrie B. Partridge, Andrew G. Truscott[*] & Randall G. Hulet

Department of Physics and Astronomy and Rice Quantum Institute, Rice University, Houston, Texas 77251, USA

[*]Present address: Research School of Physical Sciences and Engineering, Australian National University, Canberra, ACT 0200, Australia

**Attraction between atoms in a Bose-Einstein condensate renders the condensate unstable to collapse. Confinement in an atom trap, however, can stabilize the condensate for a limited number of atoms [1], as was observed with $^{7}$Li [2], but beyond this number, the condensate collapses [3-5]. Attractive condensates constrained to one-dimensional (1D) motion are predicted to form stable solitons for which the attractive interactions exactly compensate for wave packet dispersion [1]. Here we report the formation of bright solitons of $^{7}$Li atoms created in a quasi-1D optical trap. The solitons are created from a stable Bose-Einstein condensate by magnetically tuning the interactions from repulsive to attractive. A remarkable "soliton train", containing many solitons, is observed. The solitons are set in motion by offsetting the optical potential and are observed to propagate in the potential for many oscillatory cycles without spreading. Repulsive interactions between neighboring solitons are inferred from their motion.**

Dispersion and diffraction cause localized wave packets to spread as they propagate. Solitons may be formed when a nonlinear interaction produces a self-focusing of the wave packet that compensates for dispersion. Such localized structures have been observed in many physical systems including water waves, plasma waves, sound waves in liquid helium, particle physics, and in optics [6]. A Bose-Einstein condensate can be described by the nonlinear Schrödinger equation, for which the interaction term is cubic in the condensate wave function [7]. For attractive interactions, this equation has the same form as the equation for an optical wave propagating in a medium with a cubic, self-focusing (Kerr) nonlinearity and, in this sense, bright matter wave solitons in 1D are similar to optical solitons in optical fibers. Dark solitons have been recently studied in condensates with repulsive atomic interactions [8-10], but they are limited in that they can only exist within the condensate itself. Bright solitons, on the other hand, may propagate over much larger distances, and are themselves condensates. A similar experiment to ours has recently been performed by L. Khaykovich *et al.* (personal communication).

The required degree of radial confinement necessary to achieve soliton stability has been investigated theoretically [11-14]. Assuming cylindrically symmetric harmonic confinement with axial and radial oscillation frequencies of $\omega_z$ and $\omega_r$, respectively, radial excitations are suppressed in the so-called "quasi-1D" regime, where $\hbar\omega_r$ exceeds the magnitude of the mean-field interaction energy. This requirement is equivalent to a limitation on the condensate occupation number of $N < \ell_r / |a|$, where $\ell_r = (\hbar/m\omega_r)^{1/2}$ is the radial scale length, $m$ is the atomic mass, and $a$ is the $s$-wave scattering length characterizing the two-body interactions. The interactions are effectively attractive for $a < 0$ and are repulsive for $a > 0$. Achievement of the quasi-1D regime has been recently demonstrated in the case of $^7$Li [15] and $^{23}$Na [16] condensates.

The apparatus for producing a quantum degenerate gas of $^7$Li atoms was described previously [17] although a new magnetic trap with axial and radial frequencies of 70 Hz and 800 Hz, respectively, has been incorporated. Atoms in the $(F, m_F) = (2, 2)$ sublevel, where $F$ and $m_F$ are, respectively, the total angular momentum and its projection, are evaporatively cooled in the magnetic trap to a temperature of ~1 μK.

Atoms in the (1, 1) state, which are not magnetically trappable, are needed in the final stages of the experiment, so the (2, 2) atoms are transferred to an optical trap, consisting of a single, focused, red detuned laser beam propagating in the axial direction for radial confinement and a separated pair of cylindrically focused blue detuned laser beams ("end caps") propagating in the radial plane, for axial confinement. The single beam is provided by an infrared Nd:YAG laser, with a wavelength of 1064 nm, focused to a $1/e^2$ intensity radius of 47 µm and with a power of up to 750 mW. The radial confinement potential is approximately harmonic, and at the highest power, matches well with the magnetic trap potential. This single axial beam provides a very weak axial restoring potential, which is also approximately harmonic, with a frequency of ~4 Hz for oscillation amplitudes less than the Rayleigh length of 6.5 mm. The end caps are generated from the second-harmonic of another Nd:YAG laser, have $1/e^2$ radii of 22 µm axially and 100 µm radially, a power of 350 mW in each beam, and are separated by 230 µm. The end caps create a box-like potential in the axial direction, which helps to better match the magnetic trap potential. Once the optical trap lasers are switched on, the magnetic trap is switched off. After 50 ms, a bias field of up to 1000 G is applied. A period of ~200 ms is allowed for the bias field to stabilize at the chosen value before the atoms are transferred from the (2, 2) state to the (1, 1) state by an adiabatic microwave sweep of 15 ms in duration and 1 MHz in width. The purity of the (1, 1) state population is measured to be greater than 98%.

Attractive interactions between atoms in the (2, 2) state limit the number of possible condensate atoms to a very small fraction of the total number. However, interactions between (1, 1) state atoms at zero magnetic field are repulsive with a scattering length $a = 5\ a_o$, where $a_o$ is the Bohr radius [18]. Although this fact ensures the stability of the condensate, the rate of its formation is limited by the rate of thermalization, which scales as $a^2$. In order to increase this rate, a magnetically-tuned Feshbach resonance [19,20] is employed to increase the magnitude of $a$. Figure 1 shows the results of a calculation of *a* versus the magnetic field *B*, which exhibits a collisional resonance near 725 G. The rate of loss of atoms is observed to increase rapidly near the resonance (Fig. 2), in accordance with previous investigations [20,21]. Atoms are evaporatively cooled in the optical trap by halving the intensity of the infrared beam, and by tuning the bias field to 710 G, where the scattering length is large (~200 $a_o$) and positive. Absorption imaging indicates that a condensate is formed with up to ~$3 \times 10^5$ atoms. Our techniques are similar to those used to make condensates of $^{85}$Rb, where a Feshbach resonance was employed to manipulate the sign and magnitude of *a* [21].

Following the formation of the condensate at large positive *a*, the field is ramped down as $e^{-t/\tau}$ (where *t* is time and $\tau$ = 40 ms), to a selected field between 545 G and 630 G, where *a* is small and negative or small and positive (Fig. 1). The condensate can be created on the side of the optical potential by axially displacing the focus of the infrared beam relative to the centers of the magnetic trap and the box potential formed by the end caps. The end caps prevent the condensate from moving under the influence of the infrared potential until, at a certain instant, the end caps are switched off and the condensate is set in motion. The condensate is allowed to evolve for a set period of time before an image is taken. As shown in Fig. 3, the condensate spreads for *a* > 0, while for *a* < 0, non-spreading, localized structures (solitons) are formed. Solitons have been observed for times exceeding 3 s, a limitation we believe is due to loss of atoms rather than wavepacket spreading.

Remarkably, multiple-solitons ("soliton trains") are usually observed, as is evident in Figs. 3 and 4. We find that typically 4 solitons are created from an initially stationary condensate. Although multi-soliton states with alternating phase are known to be stationary states of the nonlinear Schrödinger equation [14,22,23], mechanisms for their formation are diverse. It was proposed that a soliton train could be generated by a modulational instability [24], where in the case of a condensate, the maximum rate of amplitude growth occurs at a wavelength approximately equal to the condensate healing length $\xi = (8\pi n|a|)^{-1/2}$, where $n$ is the atomic density[23]. As $a$ and $\xi$ are dynamically changing in the experiment, the expected number of solitons $N_s$ is not readily estimated from a static model. Experimentally, we detect no significant difference in $N_s$ when the time constant, $\tau$, for changing the magnetic field is varied from 25 ms to 200 ms. The dependence of $N_s$ on condensate velocity $v$ is investigated by varying the interval $\Delta t$ between the time the end caps are switched off to the time when $a$ changes sign. We find that $N_s$ increases linearly with $\Delta t$, from ~4 at $\Delta t = 0$ to ~10 at $\Delta t = 35$ ms. As the axial oscillation period is ~310 ms, $v \propto \Delta t$ in the range of $\Delta t$ investigated.

The alternating phase structure of the soliton train can be inferred from the relative motion of the solitons. Non-interacting solitons, simultaneously released from different points in a harmonic potential, would be expected to pass through one another. However, this is not observed as can be seen from Fig. 4, which shows that the spacing between the solitons *increases* near the center of oscillation and bunches at the end points. This is evidence of a short-range repulsive force between the solitons. Interaction forces between solitons have been found to vary exponentially with the distance between them and to be attractive or repulsive depending on their relative phase [25]. Because of the effect of wave interference on the kinetic energy, solitons that differ in phase by $\pi$ will repel, while those that have the same phase will attract. An alternating phase structure can be generated in the initial condensate by a phase gradient, $d\phi/dz$, across the condensate. Such a gradient may be imprinted by a condensate velocity, since $d\phi/dz = mv/\hbar$, where $m$ is the atomic mass. If $N_s$ is identified with $\phi/\pi$, the model predicts $N_s \propto v$, in agreement with the observed $v$-dependent part of $N_s$. Furthermore, for the largest $v$ and for parameters consistent with the experiment, the model gives $N_s \approx 15$, in rough agreement with observation.

6For a soliton with $a = -3\, a_o$, the calculated maximum number of atoms that ensures stability is only ~6000 per soliton [11,13], which accounts for far fewer atoms than the number contained in the initial repulsive condensate. Apparently, most of the atoms from the collapsing condensate are lost, while only a small fraction remain as solitons. Immediately after switching $a$ from positive to negative we observe a diffuse background of atoms spreading out axially. This observation is reminiscent of the jet emitted by a $^{85}$Rb condensate after switching from repulsive to attractive interactions [5]. In our system, which is in the quasi-1D regime, the remnant atoms form solitons with atom number near their stability limit.

The remarkable similarities between bright matter wave solitons and optical solitons in fibers [26] underscore the intimate connection between atom optics with Bose-Einstein condensates [27] and light optics. Many issues remain to be addressed, however, including the dynamical process of soliton formation. In addition, further investigation of soliton interactions and collisions can be undertaken with this system. Finally, we speculate that an "atomic soliton laser", based on bright matter wave solitons, may prove useful for precision measurement applications, such as atom interferometry [28].

**Acknowledgements.** We thank W. I. McAlexander for providing the coupled channels calculation, and Ben Luey for making the magnetic coils. We also thank Tom Killian and Henk Stoof for stimulating discussions. This work was supported by the U.S. National Science Foundation, the National Aeronautics and Space Administration, the Office of Naval Research, and the Welch Foundation.

**Correspondence and requests for materials should be addressed to R.G.H. (e-mail: randy@atomcool.rice.edu).**

Figure 1. Feshbach resonance. Calculation of the scattering length vs. magnetic field for atoms in the (1, 1) state of $^7$Li using the coupled channels method [18]. The field axis has been scaled in this figure by the factor 0.91, to agree with the measured resonance position of 725 G shown in Fig. 2. The scattering length is given in units of the Bohr radius.

Figure 2. Measured rate of inelastic collisional loss of atoms near the Feshbach resonance. The temperature is ~1 μK, which is above the transition temperature for Bose-Einstein condensation. The initial peak density is estimated to be ~$6 \times 10^{12}$ cm$^{-3}$. The rate of loss is given by the time for the number of trapped atoms to fall to $e^{-1}$ of the initial number. The magnetic field is determined spectroscopically by measuring the frequency of the (2, 2) → (1, 1) transition to within an uncertainty of 0.1 G.

Figure 3. Comparison of the propagation of repulsive condensates with atomic solitons. The images are obtained using destructive absorption imaging, with a probe laser detuned 27 MHz from resonance. The magnetic field is reduced to the desired value before switching off the end caps. The times given are the intervals between turning off the end caps and probing (the end caps are on for the $t$ = 0 images). The axial dimension of each image frame corresponds to 1.28 mm at the plane of the atoms. The amplitude of oscillation is ~370 μm and the period is 310 ms. The $a$ > 0 data correspond to 630 G, for which $a \approx 10$ $a_o$, and the initial condensate number is ~$3 \times 10^5$. The $a$ < 0 data correspond to 547 G, for which $a \approx -3$ $a_o$. The largest soliton signals correspond to ~5000 atoms per soliton, although significant image distortion limits the precision of number measurement. The spatial resolution of ~10 μm is significantly greater than the expected transverse dimension $\ell_r \approx 1.5$ μm.

Figure 4. Repulsive interactions between solitons. The three images show a soliton train near the two turning points and near the center of oscillation. The spacing between solitons is compressed at the turning points, and spread out at the center of the oscillation. A simple model based on strong, short-range,

repulsive forces between nearest neighbor solitons indicates that the separation between solitons oscillates at approximately twice the trap frequency, in agreement with observations. The number of solitons varies from image to image because of shot to shot experimental variations, and because of a very slow loss of soliton signal with time.  Since the axial length of a soliton is expected to vary as $1/N$ [11], solitons with small numbers of atoms produce particularly weak absorption signals, scaling as $N^2$.  Trains with missing solitons are frequently observed, but it is not clear whether this is because of a slow loss of atoms, or because of sudden loss of an individual soliton.

11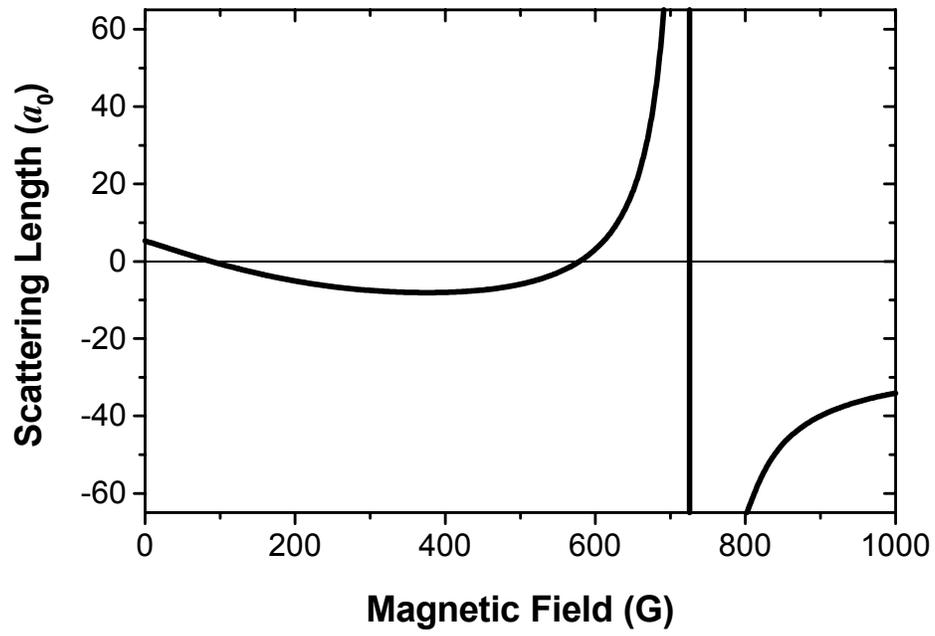

Figure #1  Strecker *et al.*

12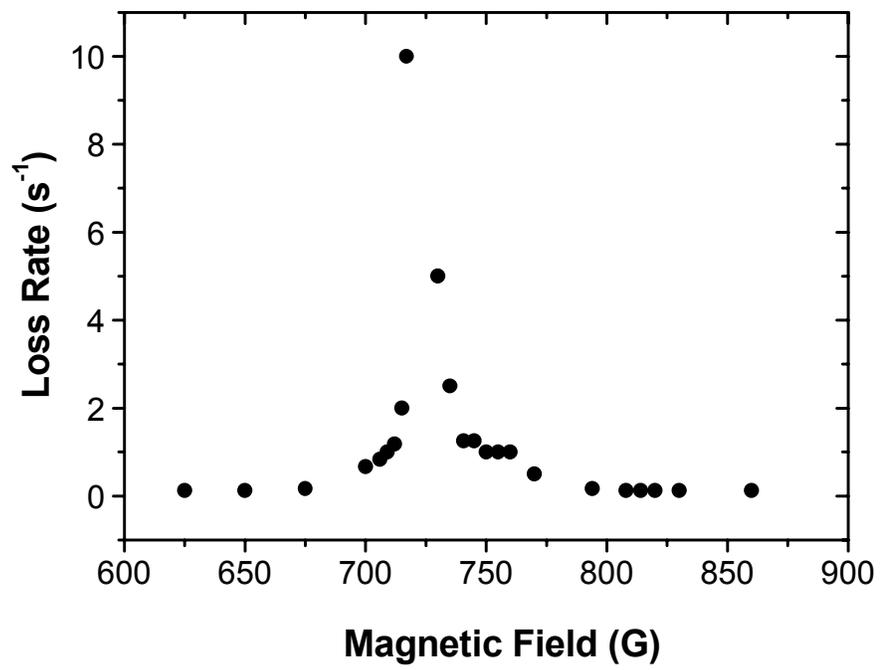

Figure #2  Strecker *et al.*

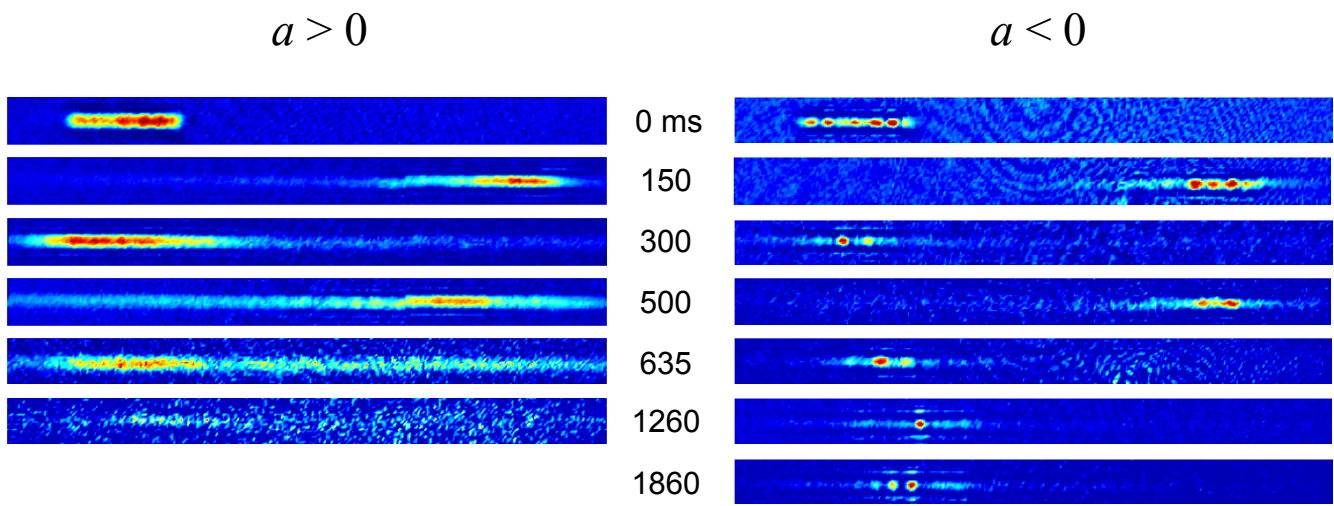

Figure 3. Strecker *et al*.

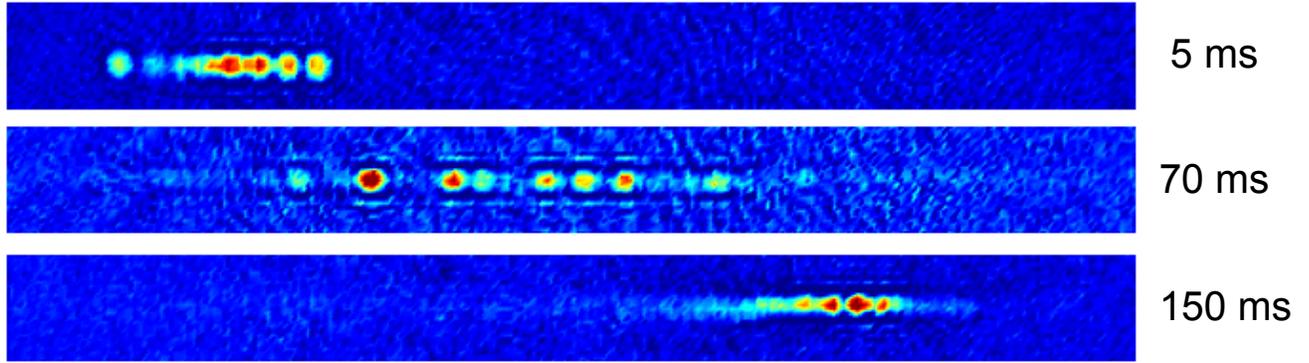

5 ms

70 ms

150 ms

Figure 4 Strecker *et al.*